\documentclass[aps,pra,floatfix,twocolumn,superscriptaddress,showpacs,10pt]{revtex4-1}
\usepackage{amssymb,amsmath,color,mciteplus,graphicx,subfigure}
\usepackage{calc,epsfig,epstopdf,color,mciteplus,bm,mathrsfs}
\usepackage{times}
\usepackage{float}
\usepackage{lipsum}
\usepackage{color}
\usepackage{amsmath}
\usepackage{ulem}
\usepackage{CJKutf8}    
\usepackage[unicode=true,
bookmarks=true,bookmarksnumbered=false,bookmarksopen=false,
breaklinks=false,pdfborder={0 0 1},backref=false,colorlinks=true]
{hyperref}
\hypersetup{linkcolor=blue, urlcolor=blue, citecolor=blue,pdfstartview={FitH}, hyperfootnotes=false, unicode=true}

\begin{document}

\title{Crosstalk-robust superconducting two-qubit geometric gates using tunable couplers}

\author{Bo-Xun Deng} 
\affiliation{Key Laboratory of Atomic and Subatomic Structure and Quantum Control (Ministry of Education), Guangdong Basic Research Center of Excellence for Structure and Fundamental Interactions of Matter, and School of Physics, South China Normal University, Guangzhou 510006, China}

\author{Jia-Qi Hu}
\affiliation{Key Laboratory of Atomic and Subatomic Structure and Quantum Control (Ministry of Education), Guangdong Basic Research Center of Excellence for Structure and Fundamental Interactions of Matter, and School of Physics, South China Normal University, Guangzhou 510006, China}

\author{Cheng-Yun Ding} 
\affiliation{School of Mathematics and Physics, Anqing Normal University, Anqing 246133, China}
\affiliation{Key Laboratory of Atomic and Subatomic Structure and Quantum Control (Ministry of Education), Guangdong Basic Research Center of Excellence for Structure and Fundamental Interactions of Matter, and School of Physics, South China Normal University, Guangzhou 510006, China}

\author{Zheng-Yuan Xue}   
\email{zyxue83@163.com}
\affiliation{Key Laboratory of Atomic and Subatomic Structure and Quantum Control (Ministry of Education), Guangdong Basic Research Center of Excellence for Structure and Fundamental Interactions of Matter, and School of Physics, South China Normal University, Guangzhou 510006, China}
\affiliation{Guangdong Provincial Key Laboratory of Quantum Engineering and Quantum Materials, \\ Guangdong-Hong Kong Joint Laboratory of Quantum Matter, and Frontier Research Institute for Physics, \\  South China Normal University, Guangzhou 510006, China}

\author{Tao Chen}   
\email{chentamail@163.com}
\affiliation{Key Laboratory of Atomic and Subatomic Structure and Quantum Control (Ministry of Education), Guangdong Basic Research Center of Excellence for Structure and Fundamental Interactions of Matter, and School of Physics, South China Normal University, Guangzhou 510006, China}
\affiliation{Guangdong Provincial Key Laboratory of Quantum Engineering and Quantum Materials, \\ Guangdong-Hong Kong Joint Laboratory of Quantum Matter, and Frontier Research Institute for Physics, \\  South China Normal University, Guangzhou 510006, China}

\date{\today}

\begin{abstract}
The design of coupler-based superconducting two-qubit gates simplifies circuit layout and alleviate frequency crowding, thereby enhancing the scalability and flexibility of quantum chips. However, in such architectures, a trade-off often exists between suppressing crosstalk and reducing gate duration, and how to achieve synergistic optimization of both remains an open challenge. To address this, this paper proposes a coupler-assisted superconducting two-qubit geometric gate scheme oriented towards crosstalk roubstness. By introducing additional parametric degrees of freedom, the scheme steers the system evolution along desired trajectories, thereby flexibly avoiding crosstalk-sensitive operational regions. Numerical simulations demonstrate that the proposed scheme can effectively suppress crosstalk errors while enabling fast gate operations, and exhibits strong robustness against typical experimental imperfections such as qubit frequency drift. Moreover, even when accounting for unavoidable high-frequency oscillation terms and qubit decoherence in realistic physical systems, our crosstalk-robust two-qubit geometric gates still achieve high fidelity. This work provides a feasible pathway toward robust and efficient two-qubit gate implementation in superconducting quantum computation.
\end{abstract}

\maketitle

\section{Introduction}

Unlike classical computers, quantum computers can simultaneously handle a vast number of computing tasks, thereby exponentially improving computational efficiency \cite{QC}. As an emerging strategy, quantum computing demonstrates the potential to address complex problems that classical computers are unable to solve. However, the current gate performance still cannot meet the requirements of fault-tolerant quantum computing, resulting in excessive resource consumption and making fault-tolerant quantum computing still face huge challenges \cite{ZP}. Therefore, enhancing the performance of quantum gates on multiple physical platforms is crucial for realizing fault-tolerant quantum computing.

Among the various quantum computing platforms, superconducting quantum circuits are recognized as one of the most promising architectures. In contemporary superconducting quantum processors, besides improving the decoherence time of qubits, reducing gate times can also fundamentally enhance the decoherence performance of the gates. However, increasing gate speed typically necessitates stronger inter-qubit coupling, which introduces more pronounced undesirable interactions, thereby creating a trade-off between gate speed and fidelity \cite{ZZ and geff}. Furthermore, since the fidelity and gate speed of two-qubit gates are currently lower than those of single-qubit gates \cite{two qubit}, this trade-off has a particularly significant impact on two-qubit gates. The primary schemes for implementing two-qubit gates include cross-resonance \cite{cross-resonance1,cross-resonance2,cross-resonance3,cross-resonance4,cross-resonance5,cross-resonance6}, parametrically tunable couplings \cite{parametrically tunable coupling0,parametrically tunable coupling1,parametrically tunable coupling2,parametrically tunable coupling3,parametrically tunable coupling4,parametrically tunable coupling5,parametrically tunable coupling6,parametrically tunable coupling7,parametrically tunable coupling8,parametrically tunable coupling9}, and tunable coupler schemes \cite{YAN,tunable coupler2,tunable coupler3,tunable coupler4,tunable coupler5,tunable coupler6}. Among these, the tunable coupler schemes can dynamically adjust the interaction between qubits and solve the problem of frequency crowding \cite{YAN}. Therefore, a substantial number of studies are currently evaluating its advantages through experimental testing.

For two-qubit gate implementations based on tunable couplers, crosstalk \cite{crosstalk review 1, crosstalk review 2} is one of the critical factors limiting the improvement of gate fidelity, with its influence persisting throughout the entire dynamical process of gate operation. Typical examples include: (i) unintended evolution of qubits caused by residual interactions \cite{tunable coupler4, CT1 1, CT1 2}; (ii) spectator-qubit effects, whereby operations on the target qubit induce errors in idle or non-participating qubits \cite{CT2 1,CT2 2}; (iii) control-line crosstalk, which is the interference of control signals on adjacent devices \cite{CT3 1}. Among these, the residual ZZ interaction, also known as ZZ crosstalk, alters the phase accumulated within the target two-qubit subspace during the execution of the two-qubit gate, resulting in an actual operation of $\text{CP}(\theta+\delta)$ rather than the intended quantum gate $\text{CP}(\theta)$ \cite{ZZ and geff,ZZ reason1,ZZ reason2}. Previous tunable coupler schemes reduce the impact of ZZ crosstalk by adjusting the coupler frequency, but this also reduces the effective coupling strength, increases the duration of gate operations, and consequently limits gate operation performance due to decoherence and error accumulation. Additionally, the academic community has proposed quantum control schemes such as dynamically decoupling \cite{DYN1,DYN2,DYN3,DYN4,DYN5,DYN6} and Hamiltonian reverse design in the rotating frame \cite{LIANG YAN} to suppress ZZ crosstalk. Although these methods effectively suppress ZZ crosstalk, they often entail increased resource consumption, introduce instability in processor performance, or add complexity to gate implementation. Therefore, achieving high-fidelity and robust two-qubit gates in a simple manner within the tunable coupler architecture remains an open challenge.

Here, we propose a coupler-assisted superconducting two-qubit geometric gate scheme oriented towards ZZ crosstalk robustness. Our approach introduces additional parametric degrees of freedom to steer the system evolution along desired trajectories, thereby flexibly avoiding crosstalk-sensitive operational regions. Furthermore, the scheme does not require complex pulse shaping or auxiliary qubit resources, thus not introducing additional control overhead. Through detailed numerical simulations, we demonstrate that the proposed scheme effectively suppresses the effect of ZZ crosstalk errors while enabling fast gate operations. It also exhibits strong robustness against typical experimental imperfections, such as qubit-frequency drift. Finally, even when accounting for unavoidable high-frequency oscillation terms and decoherence effects in realistic physical systems, our crosstalk-robust two-qubit geometric gates still achieve high fidelity.

\begin{figure}[t]
    \centering
    \includegraphics[width=1\linewidth]{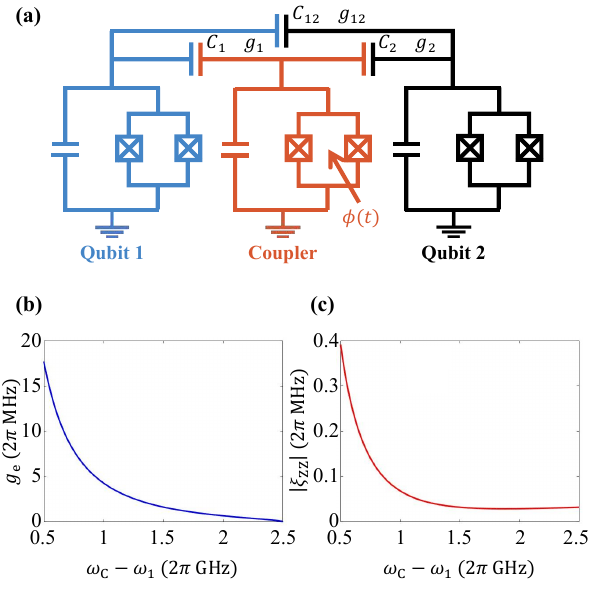}
    \caption{(a) Schematic circuit of the coupler system. Two-qubit gate is realized by applying a fast-flux bias $\phi(t)$ to the coupler SQUID loop. (b) Based on Eq. (\ref{dot1}) and Eq. (\ref{Heff}), the relationship between the effective coupling strength $g_{e}$ and $\omega_{c}-\omega_{1}$ is simulated. (c) Based on Eqs. (\ref{HZZ})-(\ref{xin}), the relationship between the absolute value of the ZZ crosstalk strength $|\xi_{ZZ}|$ and $\omega_{c}-\omega_{1}$ is simulated.}
    \label{FIG1}
\end{figure}

\section{Coupler-assisted tunable coupling}

We consider a general system consisting of two transmon qubits ($Q_1$, $Q_2$) coupled via a tunable coupler ($C$), as shown in Fig. \ref{FIG1}(a). The two qubits (the transition frequencies as $\omega_1$ and $\omega_2$) each couple to the coupler (the transition frequency as $\omega_{c}$) with a coupling strength $g_k$ ($k = 1, 2$), as well as to each other through a capacitor with a coupling strength $g_{12}$. Without loss of generality, the system original Hamiltonian can be written as
\begin{subequations}  \label{Hyuan}
\begin{align}
H_{\text{sys}}&= H_{0}+V,\\
H_{0}&=\sum_{i=1,2,c}(\omega_{i}a_{i}^{\dagger}a_{i}+\frac{\alpha_{i}}{2}a_{i}^{\dagger}a_{i}^{\dagger}a_{i}a_{i}),\\
V&= \sum_{k=1,2}g_{k}(a_{k}^{\dagger}a_{c}+a_{k}a_{c}^{\dagger})+g_{12}(a_{1}^{\dagger}a_{2}+a_{1}a_{2}^{\dagger}),
\end{align}
\end{subequations}
where $a_{i}^{\dagger}$ and $a_{i}$ ($i=1,2,c$) are, respectively, the corresponding raising and lowering operators. In the strong dispersive regime, where $g_{k} \ll |\Delta_{k}|=|\omega_{k}-\omega_{c}|$, and assuming that the coupler mode remains in its ground state, the qubit-qubit Hamiltonian with the tunable coupler decoupled can be derived by making the unitary transformation $U = \exp\{\sum_{k=1,2}{g_{k}}/{\Delta_{k}}(a_{k}^{\dagger}a_{c}-a_{k}a_{c}^{\dagger})\}$ \cite{SW1, SW2} and keeping to second order in ${g_{k}}/{\Delta_{k}}$:
\begin{align}  \label{HSW}
H=\sum_{k=1}^2(\tilde{\omega}_{k}a_{k}^{\dagger}a_{k}+\frac{\alpha_{k}}{2}a_{k}^{\dagger}a_{k}^{\dagger}a_{k}a_{k})+\tilde{g}(a_{1}^{\dagger}a_{2}+ a_{1} a_{2}^{\dagger}),
\end{align}
where 
$\tilde{\omega}_{k}=\omega_{k}+{g_{k}^{2}}/{\Delta_{k}}$ is the transformed qubit frequency, and 
$\tilde{g} = g_{12}+{g_{1}g_{2}}/{\Delta}$
is the transformed qubit-qubit coupling strength with ${1}/{\Delta}=({1}/{\Delta_{1}}+{1}/{\Delta_{2}})/2$. It is evident that the transformed coupling strength $\tilde{g}$ consists of two components, $g_{12}$ and ${g_{1}g_{2}}/{\Delta}$. When applying a flux bias $\phi(t)$ to the tunable coupler, the frequency of the tunable coupler follows the relation as \cite{omega c}
\begin{align} \label{omegac}
\omega_{c}(\phi)=\omega_{c,0}\sqrt{|\cos{(\pi\phi/\Phi_{0})}|},
\end{align}
where $\Phi_{0}$ is the flux quantum. The magnitude of ${g_{1}g_{2}}/{\Delta}$ can be adjusted via $\phi(t)$. Since the flux bias $\phi(t)$ is continuously adjustable, the  coupling strength $\tilde{g}$ can also be continuously adjusted within a certain range. To achieve parametrically tunable coupling, we apply a modulation flux on the coupler, $\phi(t)=\phi_{\text{DC}}+\phi_{\text{AC}}\cos{(\omega_{\phi}t+\varphi)}$, where $\phi_{\text{DC}}$ is the DC flux bias, and $\phi_{\text{AC}}\cos(\omega_{\phi}t+\varphi)$ is a sinusoidal fast-flux bias modulation with an amplitude $\phi_{\text{AC}}$, frequency $\omega_{\phi}$, and phase $\varphi$ \cite{parametrically tunable coupling3}. Conventional two-qubit gate construction schemes typically fix $\varphi=0$, here we introduce it as the first adjustable parameter. Expanding $\tilde{\omega}_{k}$ in the parameter $\phi_{\text{AC}}\cos{(\omega_{\phi}t+\varphi)}$ to second order under the condition $\phi_{\text{AC}}=0.1 \ll 1$ ($\phi_{\text{AC}}$ and $\phi_{\text{DC}}$ are expressed in units of $\Phi_{\text{0}}$) \cite{parametrically tunable coupling3}, we obtain
\begin{align}
\tilde{\omega}_{k}(\phi) &\approx \tilde{\omega}_{k}(\phi_{\text{DC}}) + \phi_{\text{AC}}\dfrac{\partial\tilde{\omega}_{k}}{\partial \phi} \Bigg |_{\phi\to\phi_{\text{DC}}}\cos{(\omega_{\phi}t+\varphi)}\notag\\
&\quad+\dfrac{\phi_{\text{AC}}^{2}}{2}\dfrac{\partial^{2}\tilde{\omega}_{k}}{\partial \phi^{2}} \Bigg |_{\phi\to\phi_{\text{DC}}}\cos^{2}{(\omega_{\phi}t+\varphi)}\notag\\
&=\tilde{\omega}_{k}(\phi_{\text{DC}})+\dfrac{\phi_{\text{AC}}^{2}}{4}\dfrac{\partial^{2}\tilde{\omega}_{k}}{\partial \phi^{2}} \Bigg |_{\phi\to\phi_{\text{DC}}}\notag\\
&\quad+\phi_{\text{AC}}\dfrac{\partial\tilde{\omega}_{k}}{\partial\phi}\Bigg |_{\phi\to\phi_{\text{DC}}}\cos(\omega_{\phi}t+\varphi
)\notag\\
&\quad+\dfrac{\phi_{\text{AC}}^{2}}{4}\dfrac{\partial^{2}\tilde{\omega}_{k}}{\partial\phi^{2}}\Bigg |_{\phi\to\phi_{\text{DC}}}\cos(2\omega_{\phi}t+2\varphi).
\end{align}
A similar expansion hold for $\tilde{g}$:
\begin{align}
\tilde{g}(\phi) &\approx \tilde{g}(\phi_{\text{DC}})+\dfrac{\phi_{\text{AC}}^{2}}{4}\dfrac{\partial^{2}\tilde{g}}{\partial \phi^{2}} \Bigg |_{\phi\to\phi_{\text{DC}}} \notag\\
&\quad+\phi_{\text{AC}}\dfrac{\partial\tilde{g}}{\partial\phi}\Bigg |_{\phi\to\phi_{\text{DC}}}\cos(\omega_{\phi}t+\varphi
)\notag\\
&\quad+\dfrac{\phi_{\text{AC}}^{2}}{4}\dfrac{\partial^{2}\tilde{g}}{\partial\phi^{2}}\Bigg |_{\phi\to\phi_{\text{DC}}}\cos(2\omega_{\phi}t+2\varphi).
\end{align}
Since the flux bias is often measured in units of the flux quantum $\Phi_{\text{0}}$, $\phi_{\text{AC}}\ll 1$ implies that the modulation amplitude is much smaller than $\Phi_{\text{0}}$, ensuring that in the expansion, we can effectively ignore the influence of higher-order terms. According to the definition of $\tilde{g}$, the first-order partial derivative expansion of $\tilde{g}$ with respect to $\phi$ is
\begin{align}   \label{dot1}
\dfrac{\partial\tilde{g}}{\partial\phi}=-\dfrac{g_{1}g_{2}\dot{\Delta}(\phi)}{\Delta^{2}(\phi)}=-\dfrac{2g_{1}g_{2}(\dot{\Delta}_{1}\Delta_{2}^{2}+\Delta_{1}^{2}\dot{\Delta}_{2})}{\Delta^{2}(\phi)(\Delta_{1}+\Delta_{2})},
\end{align}
where
\begin{align}   \label{dotDelta}
\dot{\Delta}_{k}=-\dot{\omega}_{c}=\dfrac{1}{2}\dfrac{\pi}{\Phi_{0}}\omega_{c,0}(\sin{(\pi\dfrac{\phi}{\Phi_{0}})}/\sqrt{\cos{(\pi\dfrac{\phi}{\Phi_{0}})}}).
\end{align}
Once $\omega_{c}$ is determined, $\phi$ is also consequently determined; therefore, at this point, the values of $\dot{\Delta}_{k}$ and $\frac{\partial\tilde{g}}{\partial\phi}$ can be determined respectively.

In the interaction frame, the oscillating $a_{k}^{\dagger}a_{k}$ terms appear as a trigonometric function within the argument of the natural exponential function. Since these terms average to zero over time, their net contribution to the system dynamics can be effectively neglected. After updating Eq. (\ref{HSW}) to include all other expansion terms, the Hamiltonian in a frame rotating at the qubit frequency becomes (with all quantities evaluated at $\phi=\phi_{\text{DC}}$)
\begin{align}
H' &= g' \big[e^{i\Delta_{12,\phi}t}|10\rangle\langle01|+e^{i(\Delta_{12,\phi}-\alpha_{2})t}\sqrt{2}|11\rangle\langle02|\notag\\
&\quad+e^{i(\Delta_{12,\phi}+\alpha_{1})t}\sqrt{2}|20\rangle\langle11|+\text{H.c.}\big],
\end{align}
where 
\begin{align}
g'&= (\tilde{g}+\frac{\phi_{\text{AC}}^{2}}{4}\frac{\partial^{2}\tilde{g}}{\partial\phi^{2}})+\phi_{\text{AC}}\frac{\partial\tilde{g}}{\partial\phi}\cos(\omega_{\phi}t+\varphi)\notag\\
&\quad\,\,\,+\frac{\phi_{\text{AC}}^{2}}{4}\frac{\partial^{2}\tilde{g}}{\partial\phi^{2}}\cos(2\omega_{\phi}t+2\varphi),
\end{align}
the state vector $|nm\rangle=|n\rangle_{Q_{1}}\otimes|m\rangle_{Q_{2}}$ with $n,m \in \text{N}$, and 
$\Delta_{12,\phi}=\tilde{\omega}_{1}(\phi_{\text{DC}})-\tilde{\omega}_{2}(\phi_{\text{DC}})+\frac{\phi_{\text{AC}}^{2}}{4}(\frac{\partial^{2}\tilde{\omega}_{1}}{\partial\phi^{2}}-\frac{\partial^{2}\tilde{\omega}_{2}}{\partial\phi^{2}})$.

In the previous scheme \cite{ZZ and geff}, the two-qubit $i$SWAP gates were realized by applying a sinusoidal fast-flux bias modulation pulse to the tunable coupler, with the frequency set to $\omega_{\phi}=\Delta_{12,\phi}$ and the phase fixed at $\varphi=0$. In addition, it was found that increasing the coupler frequency $\omega_c$ can partially suppress ZZ crosstalk. However, this approach simultaneously extends the gate duration, thereby increasing the exposure of the system to other error sources, such as decoherence and high-frequency oscillatory terms. As a result, despite achieving partial suppression of ZZ crosstalk, the overall gate fidelity is ultimately reduced due to the accumulation of these additional errors. The $i$SWAP gate can be realized by setting $\omega_{\phi}=\Delta_{12,\phi}$, which allows the effective Hamiltonian’s subspace to be $\{|10\rangle,|01\rangle\}$; whereas the CZ gate can be achieved by setting $\omega_{\phi}=\Delta_{12,\phi}-\alpha_{2}$, resulting in the effective Hamiltonian subspace being $\{|11\rangle,|02\rangle\}$. To provide additional parameter for CZ gate performance optimization, we introduce the second adjustable parameter $\delta$ by setting $\omega_{\phi}=\Delta_{12,\phi}-\alpha_{2}+\delta$ with $\delta \ll \Delta_{12,\phi}-\alpha_{2}$. By replacing $\Delta_{12,\phi}$ with $\omega_{\phi}+\alpha_{2}-\delta$ and using Euler's formula, we get
\begin{align}   \label{Hhigh-order}
H'' &=\frac{\phi_{\text{AC}}}{2}\frac{\partial\tilde{g}}{\partial\phi}[e^{-i(\delta t+\varphi)}\sqrt{2}|11\rangle\langle02|+\text{H.c.}]\notag\\
&\quad+\frac{\phi_{\text{AC}}}{2}\frac{\partial\tilde{g}}{\partial\phi}[e^{i(2\omega_{\phi}t-\delta t+\varphi)}\sqrt{2}|11\rangle\langle02|+\text{H.c.}]\notag\\
&\quad+\frac{\phi_{\text{AC}}^{2}}{8}\frac{\partial^{2}\tilde{g}}{\partial\phi^{2}}[e^{-i(\omega_{\phi} t+\delta t+2\varphi)}\sqrt{2}|11\rangle\langle02|+\text{H.c.}]\notag\\
&\quad+\frac{\phi_{\text{AC}}^{2}}{8}\frac{\partial^{2}\tilde{g}}{\partial\phi^{2}}[e^{i(3\omega_{\phi}t-\delta t+2\varphi)}\sqrt{2}|11\rangle\langle02|+\text{H.c.}]\notag\\
&\quad+(\tilde{g}+\frac{\phi_{\text{AC}}^{2}}{4}\frac{\partial^{2}\tilde{g}}{\partial\phi^{2}})[e^{i(\omega_{\phi}t-\delta t)}\sqrt{2}|11\rangle\langle02|+\text{H.c.}]\notag\\
&\quad+\cdots . 
\end{align}
Given that $\delta \ll \Delta_{12,\phi}-\alpha_{2}$, all terms except the first one is treated as high-frequency oscillation terms and are consequently neglected. To enable the effective Hamiltonian with a more general form and simultaneously provide additional tunable parameter freedom for subsequent trajectory optimization strategy, we apply the unitary transformation $V(t)=\exp(-i{\Delta_{\text{e}}}(|11\rangle\langle11|-|02\rangle\langle02|)t/2)$ which transforms the system from the original interaction picture to one characterized by an effective detuning $\Delta_{e}$. This transformation decomposes the phase $\delta$ that originally appears in the coupling term into two parts: one part remains in the effective coupling term, together with the initial phase of the pulse $\varphi$, forming the phase factor $e^{-i\phi_{\text{e}}(t)}$; the other part manifests as an effective detuning term $\Delta_{e}\tilde{\sigma}_{z}$ that acts on the $\{|11\rangle, |02\rangle\}$ subspace. Consequently, the explicit form of the effective Hamiltonian is
\begin{align} \label{Heff}
H_{\text{e}} &= -\frac{1}{2}\Delta_{\text{e}}\tilde{\sigma}_{z}+\frac{g_{\text{e}}}{2}[e^{-i\phi_{\text{e}}}|11\rangle \langle02|+\text{H.c.}],
\end{align}
where $g_{\text{e}} =\sqrt{2}{\phi_{\text{AC}}}\frac{\partial\tilde{g}}{\partial\phi}$\, $\phi_{\text{e}} = (\delta-\Delta_{\text{e}}) t + \varphi$, and $\tilde{\sigma}_{z} = |11\rangle\langle11|-|02\rangle\langle02|$. Therefore, once the magnitudes of $\phi_{\text{DC}}$ and $\phi_{\text{AC}}$ are determined, the effective coupling strength $g_{\text{e}}$ becomes a constant. Meanwhile, by adjusting the parameters $\delta$ and $\varphi$, the specific form of the effective Hamiltonian can be modified. From Eq. (\ref{dot1}) and Eq. (\ref{dotDelta}), it can be seen that changing $\omega_{c}$ also affects the magnitude of $\frac{\partial\tilde{g}}{\partial\phi}$, thereby influencing the value of $g_{e}$. To further investigate the relationship between $\omega_{c}$ and $g_{e}$, we used numerical simulations to generate Fig. \ref{FIG1}(b). It is evident that the effective coupling strength $g_{e}$ decreases with the increase of $\omega_{c}$. Here, we set a conservative parameter region with the detuning of $\Delta_{12}/2\pi=0.5\;\text{GHz}$, coupler frequency $\omega_{c,0}/2\pi=7.5\; \text{GHz}$, qubit anharmonicity $\alpha_{1}/2\pi=\alpha_{2}/2\pi=-200\;\text{MHz}$. Additionally, specify the nearest-neighbor coupling strength $g_{1}/2\pi=g_{2}/2\pi=86\;\text{MHz}$ and next-nearest-neighbor coupling strength $g_{12}/2\pi=5\;\text{MHz}$, in accordance with state-of-the-art technologies of superconducting qubits \cite{P2,P3}.

Our objective is to effectively suppress various sources of errors, including ZZ crosstalk, which arises from residual interactions between qubits, can induce spurious phase accumulation during gate operations, thereby reducing gate fidelity. To quantitatively assess the impact of ZZ crosstalk on gate fidelity, here we define
\begin{align}   \label{HZZ}
H_{ZZ}={\xi_{ZZ}}\sigma_{Z}^{1}\otimes\sigma_{Z}^{2},
\end{align}
where $\xi_{ZZ}$ denotes the ZZ crosstalk strength, with $\sigma_{Z}^{1}$ and $\sigma_{Z}^{2}$ representing the corresponding Pauli-Z operators. The conventional Schrieffer-Wolff (SW) transformation method constructs an effective Hamiltonian through second-order truncation. However, to accurately calculate the ZZ crosstalk between $Q_{1}$ and $Q_{2}$, it is not enough to only keep up to the second order in the SW transformation. Therefore, we use the perturbation approach \cite{perturbation} to derive the ZZ crosstalk coupling to second, third and fourth order of the system original Hamiltonian Eq. \ref{Hyuan}. The result for the ZZ crosstalk strength $\xi_{ZZ}$, can be defined as 
$\xi_{ZZ} = \xi^{(0)}+\xi^{(1)}+\xi^{(2)}+\xi^{(3)}+\xi^{(4)}$,
where $\xi^{(z)}$ denotes the $z$th-order ZZ crosstalk strength, defined as $\xi^{(z)} \equiv E_{11}^{(z)}-E_{10}^{(z)}-E_{01}^{(z)}+E_{00}^{(z)}$ with 
\begin{subequations}   \label{Energyn}
\begin{align}
E_{s}^{(0)}&=H_{s}^{(0)}, E_{s}^{(1)}=V_{ss},\\
E_{s}^{(2)}&=\sum_{j\ne s}\dfrac{|V_{sj}|^{2}}{E_{sj}},\\
E_{s}^{(3)}&=\sum_{j, h\ne s}\dfrac{V_{sj}V_{jh}V_{hs}}{E_{sj}E_{sh}},\\
E_{s}^{(4)}&=\sum_{j, h, l\ne s}\dfrac{V_{sj}V_{jh}V_{hl}V_{ls}}{E_{sj}E_{sh}E_{sl}}+\sum_{j, h\ne s}\dfrac{|V_{sj}|^{2}|V_{sh}|^{2}}{E_{sj}^{2}E_{sh}},
\end{align}
\end{subequations}
where $H_{s}^{(0)}=\langle s |H_{0}| s\rangle$, $V_{sj}= \langle s |V| j\rangle$, $E_{sj}=E_{s}^{(0)}-E_{j}^{(0)}$. The indices $j,\,h,\,l$ correspond to the state $|Q_{1}\,Q_{2},\,C\rangle $ in the set $\{|00,\,0\rangle, |01,\,0\rangle, |10,\,0\rangle, |11,\,0\rangle, |01,\,1\rangle, |10,\,1\rangle, |02,\,0\rangle,\\ |20,\,0\rangle,\, |00,\,2\rangle\}$, while $s$ belongs to the set $\{\,|00,\,0\rangle,\, |01,\,0\rangle,\\ |10,\,0\rangle, |11,\,0\rangle\}$. Thus, according to Eq. (\ref{Energyn}), and after making the approximation, we have
\begin{subequations}  \label{xin}
\begin{align}
\xi^{(0)}& =
\xi^{(1)}=0,\xi^{(2)}=\frac{2g_{12}^{2}(\alpha_{1}+\alpha_{2})}{(\Delta_{12}+\alpha_{1})(\Delta_{12}-\alpha_{2})},\\
\xi^{(3)}&=2g_{12}g_{1}g_{2} \bigg[ \frac{1}{\Delta_{1}}\left(\frac{2}{\Delta_{12}-\alpha_{2}}-\frac{1}{\Delta_{12}}\right)\notag\\
&\quad -\frac{1}{\Delta_{2}}\left(\frac{2}{\Delta_{12}+\alpha_{1}}-\frac{1}{\Delta_{12}}\right) \bigg],\\
\xi^{(4)}&= \frac{2g_{1}^{2}g_{2}^{2}}{\Delta_{1}+\Delta_{2}-\alpha_{c}}\left[ \frac{1}{\Delta_{1}}+\frac{1}{\Delta_{2}}\right]^{2}\notag\\
&\quad +\frac{g_{1}^{2}g_{2}^{2}}{\Delta_{1}^{2}}\left[ \frac{2}{\Delta_{12}-\alpha_{2}}-\frac{1}{\Delta_{12}}-\frac{1}{\Delta_{2}}\right]\notag\\
&\quad +\frac{g_{1}^{2}g_{2}^{2}}{\Delta_{2}^{2}}\left[ -\frac{2}{\Delta_{12}+\alpha_{1}}+\frac{1}{\Delta_{12}}-\frac{1}{\Delta_{1}}\right].
\end{align}
\end{subequations}
Based on Eq. (\ref{xin}), we can clearly observe the relationship between the ZZ crosstalk strength $\xi_{ZZ}$ and the coupler frequency $\omega_{c}$. Therefore, we plotted Fig. \ref{FIG1}(c) using Eqs. (\ref{HZZ})-(\ref{xin}) in conjunction with the system parameters described earlier. From Fig. \ref{FIG1}(b) and Fig. \ref{FIG1}(c), it is not difficult to observe that the effective coupling strength $g_{\text{e}}$ and the crosstalk strength $\xi_{ZZ}$ both decrease with the increase in the coupler frequency. Therefore, while conventional methods can reduce crosstalk strength by tuning the coupler frequency, these methods simultaneously weakens the effective coupling strength between qubits, thereby prolonging gate time and increasing decoherence. 

The introduction of adjustable parameters $\delta$ and $\varphi$ through magnetic flux modulation is significant for addressing the dual challenges of ZZ crosstalk and decoherence. By utilizing these parameters, we can enhance optimal control techniques, thereby overcoming the limitation of previous methods that lacked parameter adjustability and were confined to a fixed evolution trajectory, and ultimately identifying the optimal evolution trajectory under the combined constraints of ZZ crosstalk and decoherence.

\section{Trajectory optimization for crosstalk suppression}

\begin{figure}[t]
    \centering
    \includegraphics[width=1\linewidth]{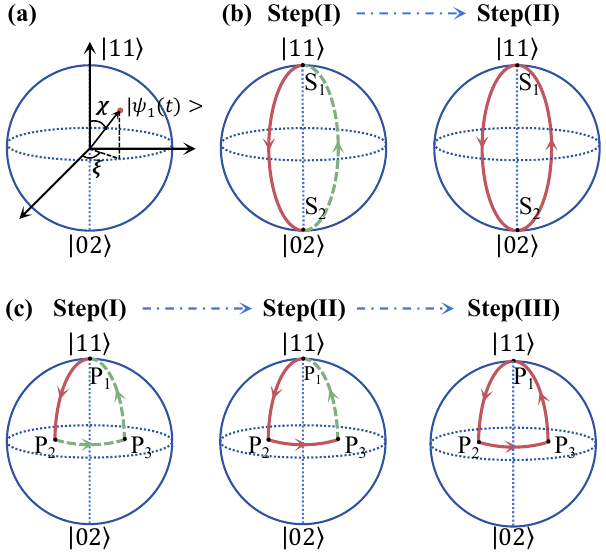}
    \caption{(a) The position coordinates of state vector $|\psi_{1}(t)$ at a certain moment on the Block sphere. (b) The evolution details of conventional single-loop nonadiabatic geometric trajectory.
    (c) The evolution details of our unconventional nonadiabatic geometric trajectory on the Block sphere, where $\chi$ and $\xi$ can be adjusted arbitrarily according to the trajectory requirements.}
    \label{Block}
\end{figure}

To obtain the optimal evolution trajectory for suppress ZZ crosstalk, we propose a trajectory correction scheme based on geometric phase. We now consider the Hamiltonian as show in Eq. (\ref{Heff}) and choose a set of dressed states, that are orthogonal to each other, as
\begin{equation}
|\Psi_{1}(t)\rangle = e^{if_{1}(t)}|\psi_{1}(t)\rangle,\quad 
|\Psi_{2}(t)\rangle = e^{if_{2}(t)}|\psi_{2}(t)\rangle,
\end{equation}
where
\begin{subequations}
\begin{align}
|\psi_{1}(t)\rangle&=\cos\frac{\chi(t)}{2}|11\rangle+\sin\frac{\chi(t)}{2}e^{i\xi(t)}|02\rangle,\\\
|\psi_{2}(t)\rangle&=\sin\frac{\chi(t)}{2}e^{-i\xi(t)}|11\rangle-\cos\frac{\chi(t)}{2}|02\rangle.
\end{align}
\end{subequations}
Besides, to satisfy the boundary condition of cyclic evolution
\begin{align}
|\psi_{k}(0)\rangle=|\psi_{k}(\tau)\rangle,
\end{align}
we need to ensure $\chi(0)=\chi(\tau)=0$, and this implies that $|\psi_{2}(0)\rangle$ is not in the computational subspace. Therefore, we take the $|\psi_{1}(t)\rangle$ as an example, on which the overall phase $f_{1}(t)$ is accumulated at the final time. The detail evolution trajectory of $|\psi_{1}(t)\rangle$ is shown on the block sphere in Fig. \ref{Block} by visualized parameters $\chi(t)$ and $\xi(t)$, which represent the polar and azimuth angles, respectively, in the range of $[0, \pi]$ and $[0, 2\pi)$. Meanwhile, by solving the Schr{\"o}dinger equation, the parameter-limited relationships can be confirmed as
\begin{subequations}   \label{chiandxi}
\begin{align}
\dot{\chi}(t) &= g_{\text{e}}\sin[\phi_{\text{e}}(t)-\xi(t)],\\\
\dot{\xi}(t) &= -\Delta_{\text{e}}(t)-g_{\text{e}}\cot\chi(t)\cos[\phi_{\text{e}}(t)-\xi(t)].
\end{align}
\end{subequations}

In addition, the overall phase accumulated during the evolution period $\tau$ is written as
\begin{align}
\gamma=f_{1}(\tau)=-f_{2}(\tau)=\int_{0}^{\tau}\frac{\dot{\xi}(t)[1-\cos\chi(t)]+\Delta_{e}(t)
}{2\cos\chi(t)}dt,
\end{align}
which includes two parts, namely, the dynamical phase
\begin{align}
\gamma_{d}&=-\int_{0}^{\tau}\langle\Psi_{1}(t)|H_{e}(t)|\Psi_{1}(t)\rangle dt\notag\\
&=\frac{1}{2}\int_{0}^{\tau}\frac{\dot{\xi}(t)\sin^{2}\chi(t)+\Delta_{e}(t)}{\cos\chi(t)}dt,
\end{align}
and the geometric phase
\begin{align}
\gamma_{g}=\gamma-\gamma_{d}=-\frac{1}{2}\int_{0}^{\tau}\dot{\xi}(t)[1-\cos\chi(t)] dt,
\end{align}
where $\gamma_{g}$ is the geometric pahse \cite{geometric phase,wang2001,zhu2002,xuereview}, as it is given by half of the solid angle enclosed by the evolution trajectory.

Therefore, after determining $\chi(t), \xi(t)$ and $\gamma$, we can ultimately express the evolution operator as
\begin{align}
U(\gamma) &= |\Psi_{1}(\tau)\rangle\langle\Psi_{1}(0)|+|\Psi_{2}(\tau)\rangle\langle\Psi_{2}(0)|\notag\\
&= e^{i\gamma}|\psi_{1}(\tau)\rangle\langle\psi_{1}(0)|+e^{-i\gamma}|\psi_{2}(\tau)\rangle\langle\psi_{2}(0)|\notag\\
&= e^{i\gamma}|11\rangle\langle11|+e^{-i \gamma}|02\rangle\langle02|.
\end{align}
Once the evolution operator is established, the time-dependent control parameters $g_{\text{e}}(t)$, $\phi_{\text{e}}(t)$, and $\Delta_{\text{e}}(t)$ directly dictate the evolution details of state vector $|\psi_{k}(t)\rangle$, which is characterized by $\xi(t)$ and $\chi(t)$. This indicates that by adjusting the Hamiltonian parameters, arbitrary desired evolution trajectories can be determined.

Consider the simplest evolution trajectory depicted in Fig. \ref{Block}(b), which evolves strictly along the longitude of the Bloch sphere and is referred to as the conventional single-loop nonadiabatic geometric computation scheme (SNGQC) \cite{SNGQC1,SNGQC2,liangyan,SNGQC3}. This scheme is commonly employed in experiments to implement a controlled-phase gate. It strictly eliminates the dynamical phase, starting from $\text{S}_{1}$ and going along $\text{S}_{1} \to \text{S}_{2} \to \text{S}_{1}$. Specifically, in the coordinates $(\chi, \xi)$, the detailed steps are follows: (I) Firstly, the state $|\psi_{1}(t)\rangle$ begins at the North Pole $\text{S}_{1}(0, \xi_{1})$ and evolves along the line of longitude at $\xi(t)=\xi_{1}$, reaching the South Pole $\text{S}_{2}(\pi, \xi_{1})$ at time $t = \tau_{1}$. Subsequently, the azimuthal angle is adjusted to $\xi(\tau_{1}+\epsilon)=\xi_{2}$ within an infinitesimal duration $\epsilon \ll 0$. (II) Next, the state $|\psi_{1}(t)\rangle$ returns to the North Pole $\text{S}_{1}(0, \xi_{2})$ by following the line of longitude at $\xi(t)=\xi_{2}$.

\begin{figure*}[t]
    \centering
    \includegraphics[width=1\linewidth]{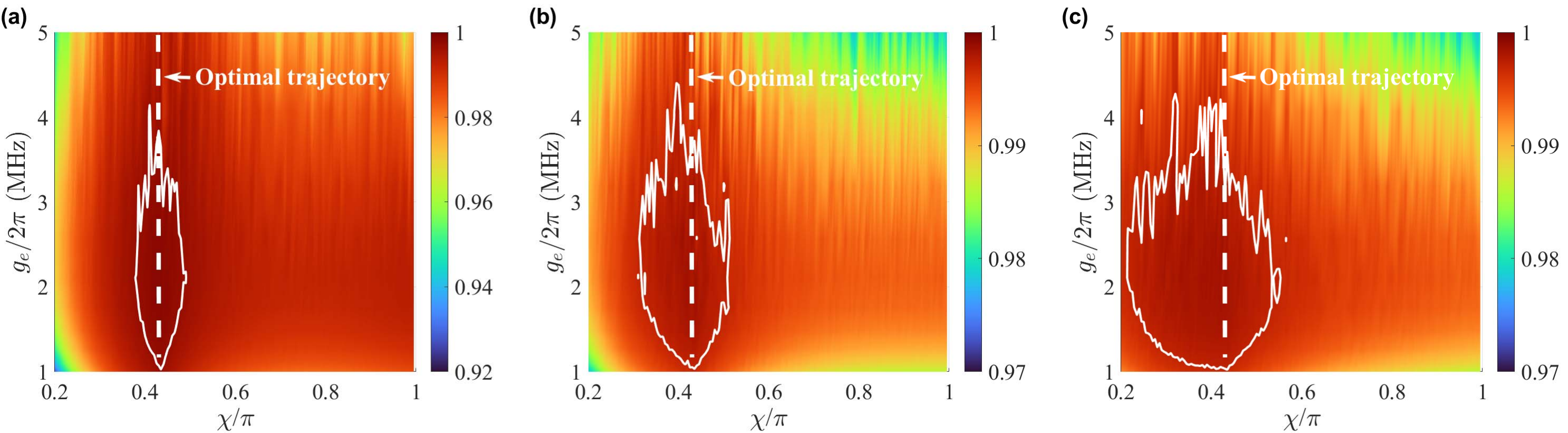}
    \caption{The fidelity of (a) $U_{\text{T}}^{\text{UG}}(\pi)$, (b) $U_{\text{T}}^{\text{UG}}(\pi/2)$ and (c) $U_{\text{T}}^{\text{UG}}(\pi/4)$ gates as functions of $\chi$ and $g_{\text{e}}$, respectively, with a fidelity contour of 0.997 plotted in each figure.}
    \label{Path}
\end{figure*}

According to the parameter relationship in Eq. (\ref{chiandxi}), we can determine the Hamiltonian parameters as
\begin{align}
&\ t\in[0,\tau_{1}):g_{\text{e}}\tau_{1} =\chi=\pi,
\ \ \phi_{\text{e}}(t) = \xi_{1}+\frac{\pi}{2}, \ \ \Delta_{e}=0,\notag\\
&\ t\in[\tau_{1},\tau]:g_{\text{e}}(\tau-\tau_{1}) =\chi, \ \
\phi_{\text{e}}(t) = \xi_{2}-\frac{\pi}{2}, \ \ \Delta_{e}=0.
\end{align}
However, the starting point $(\chi_{0}, \xi_{0})$ of the evolution process and geometric phase $\gamma_{g}$ are determined when constructing a specific quantum gate. This scheme offers only one fixed evolution trajectory, which limits the degrees of freedom for enhancing the crosstalk-resistant feature of geometric gates. 

Therefore, it is necessary to develop an unconventional nonadiabatic geometric quantum computation scheme (UNGQC) by incorporating latitude evolution trajectory. In contrast, UNGQC requires only that the dynamic phase $\gamma_{d}$ maintains a fixed proportional relationship with the geometric phase $\gamma_{g}$, denoted as
$\gamma_{d}=\eta\gamma_{g}$ with $\eta\ne 0, 1$,
where the proportionality coefficient $\eta$ is maintained at a constant value. Therefore, based on the the previously single-loop evolution trajectory, we construct the triangle-cap evolution trajectory illustrated in Fig. \ref{Block}(c), proceeding along $\text{P}_{1}\;\to\;\text{P}_{2}\;\to\;\text{P}_{3}\;\to\;\text{P}_{1}$. In the coordinates $(\chi, \xi)$, the detailed steps are as follows: (I) Firstly, the state $|\psi_{1}(t)\rangle$ begins at the north pole $\text{P}_{1}(0,\xi_{1})$ and evolves along the longitude line with $\xi(t)=\xi_{1}$ to the point $\text{P}_{2}(\chi,\xi_{1})$ at time $\tau_{1}$. (II) Then, the state $|\psi_{1}(t)\rangle$ evolves along the latitude line with $\chi(t)=\chi$ to the point $\text{P}_{3}(\chi,\xi_{2})$ at time $\tau_{2}$. (III) Finally, the state $|\psi_{1}(t)\rangle$ returns to the north pole $\text{P}_{1}(0,\xi_{2})$ at the final time $\tau$ along the longitude line with $\xi(t)=\xi_{2}$.

Consequently, based on the parameter relationship in Eq. (\ref{chiandxi}), we can determine the Hamiltonian parameters $g_{\text{e}}(t)$ and $\phi_{\text{e}}(t)$ corresponding to these three trajectory segments $t \in [0,\tau_{1})$, $[\tau_{1},\tau_{2})$, and $[\tau_{2},\tau]$ as follows
\begin{subequations}  \label{pathparameter}
\begin{flalign}      
&\ t\in[0,\tau_{1}):g_{\text{e}}\tau_{1} = \chi,
\;\phi_{\text{e}}(t) = \xi_{1}+\frac{\pi}{2}, \; \Delta_{e}=0,&\\
&\ t\in[\tau_{1},\tau_{2}):g_{\text{e}}(\tau_{2}-\tau_{1})=2\gamma\sin{\chi}\frac{ [(1+\eta)\cos\chi-\eta]}{(1+\eta)(1-\cos\chi)},&\notag\\
&\ \quad\quad\quad\quad\phi_{\text{e}}(t) = \xi_{1}-\frac{g_{\text{e}}(t-\tau_{1})}{\sin\chi[(1+\eta)\cos\chi-\eta]},&\notag\\
&\ \quad\quad\quad\quad \Delta_{e}=\dfrac{g_{\text{e}}}{\sin{\chi}[(1+\eta)\cos{\chi}-\eta]}-\dfrac{g_{\text{e}}}{\tan\chi},\,&\\
&\ t\in[\tau_{2},\tau]:g_{\text{e}}(\tau-\tau_{2}) =\chi,
\;\phi_{\text{e}}(t) = \xi_{2}-\frac{\pi}{2}, \;\Delta_{e}=0.&
\end{flalign}
\end{subequations}
Different geometric trajectories can be obtained by varying $\chi$ within the range of $[0,\pi]$. When $\chi=\pi$, the UNGQC scheme will be equivalent to the SNGQC scheme. Therefore, we are no longer confined to a specific evolution trajectory; instead, we can select the most satisfactory evolution trajectory based on the distinct characteristics of different trajectories.

\begin{figure*}[t]
\centering
    \includegraphics[width=1\linewidth]{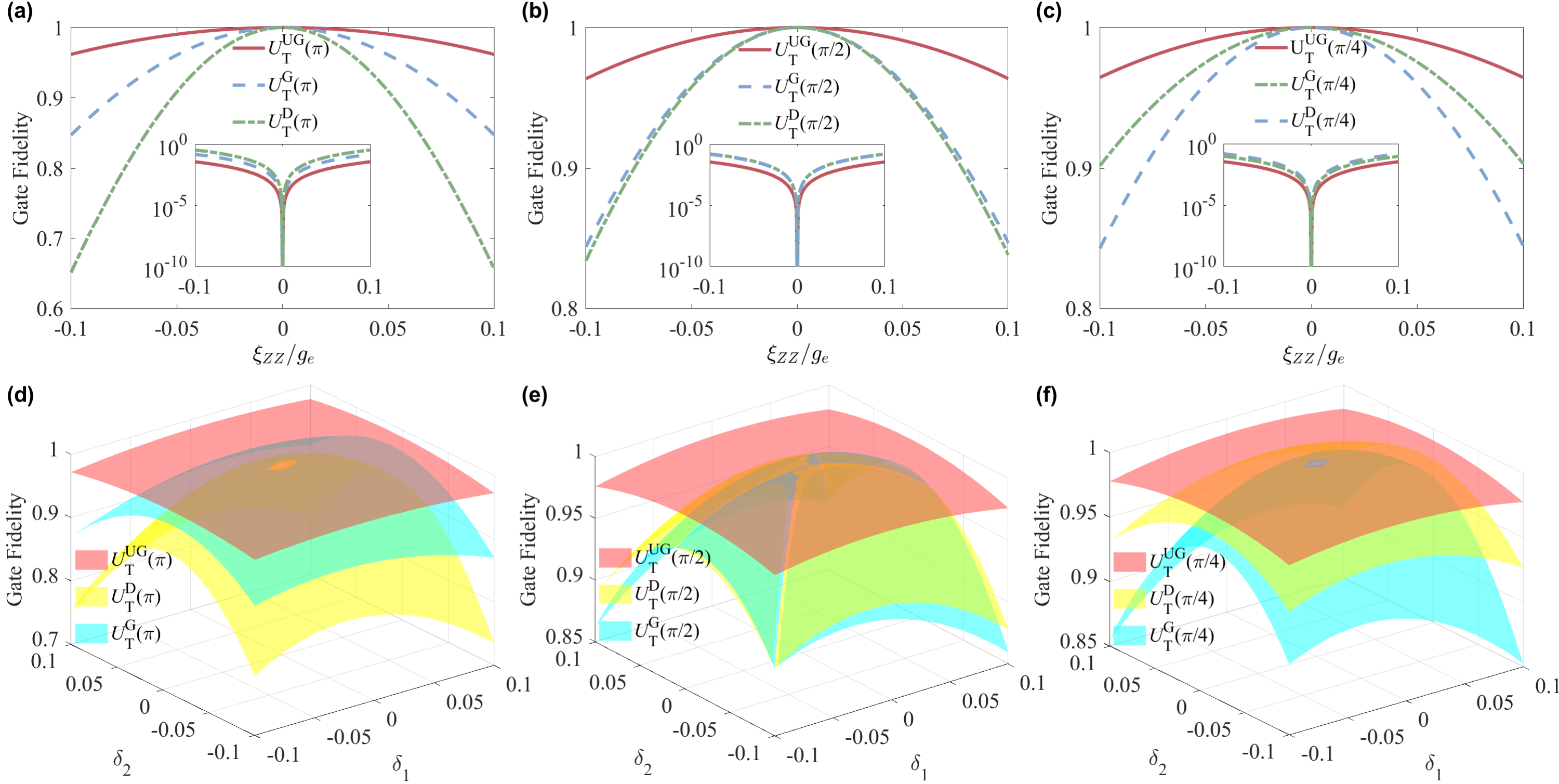}
    \caption{Utilize the unconventional nonadiabatic geometric trajectory, that determined by the optimal trajectory of $\chi=0.43\;\pi$ with the low crosstalk sensitivity, to implement the robust geometric (a) $U_{\text{T}}(\pi)$, (b) $U_{\text{T}}(\pi/2)$, (c) $U_{\text{T}}(\pi/4)$ gates. The vertical axis of subfigures in the figure represents the gate infidelity (i.e., $1-F$). In the same way, (d) $U_{\text{T}}(\pi)$, (e) $U_{\text{T}}(\pi/2)$, (f) $U_{\text{T}}(\pi/4)$ gates (red surface) can also demonstrate a stronger suppression on qubit frequency drift error than conventional geometric and dynamical counterparts.}
    \label{Robust}
\end{figure*}

\begin{figure*}[t]
    \centering
    \includegraphics[width=1\linewidth]{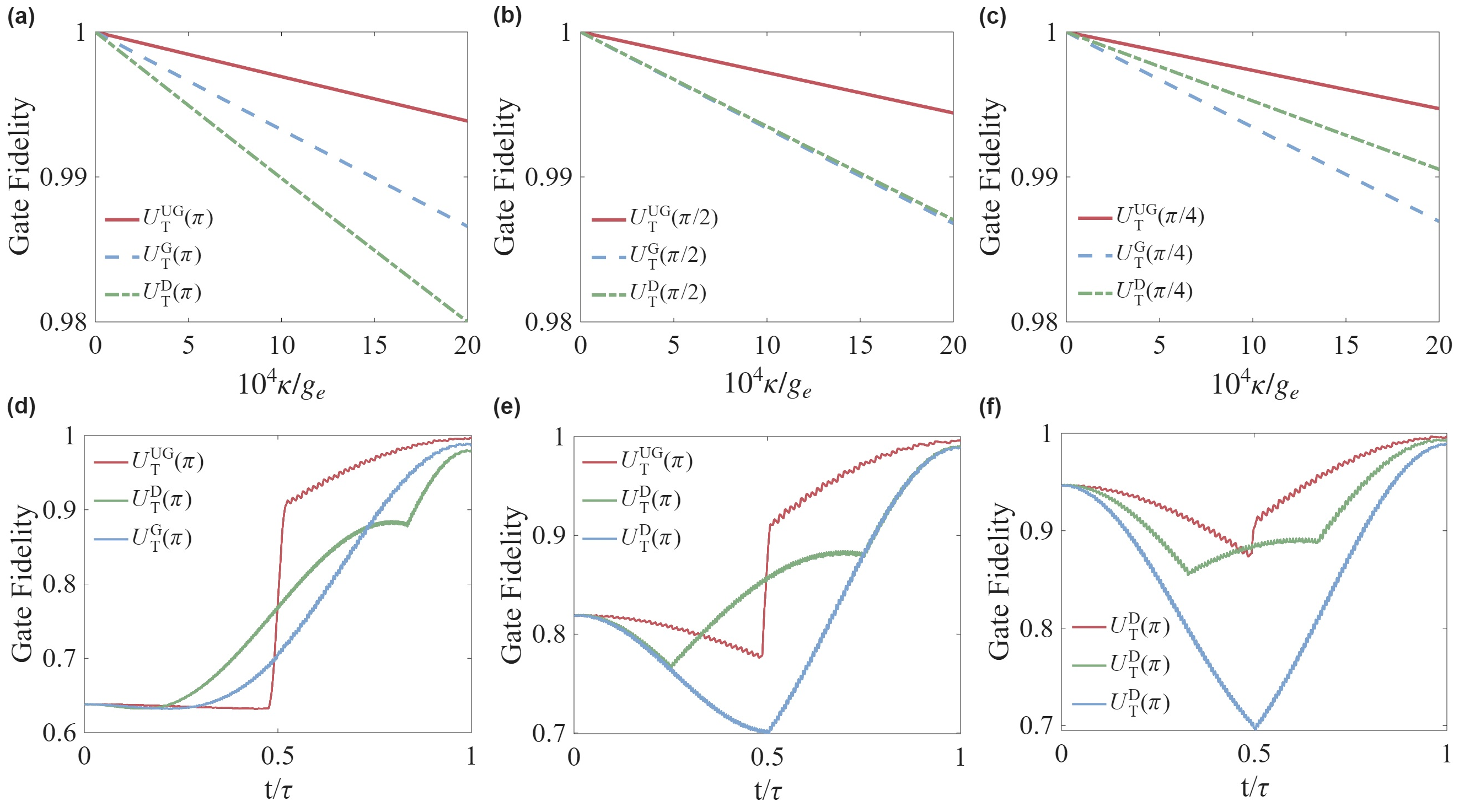}
    \caption{Comparison of performance under decoherence in different schemes for (a) $U_{\text{T}}(\pi)$, (b) $U_{\text{T}}(\pi/2)$ and (c) $U_{\text{T}}(\pi/4)$ gates. And we further conducted individual analyses of the gate fidelity under the combined influence of crosstalk, high-frequency oscillation and decoherence for (d) $U_{\text{T}}(\pi)$, (e) $U_{\text{T}}(\pi/2)$ and (f) $U_{\text{T}}(\pi/4)$ gates.}
    \label{Decoherence}
\end{figure*}

Here, we define the gate fidelity $F$ with ZZ crosstalk and high-frequency oscillation error as \cite{F1, F2,F3}
\begin{align}
F=\text{Tr}(U^{\dagger}_{ideal}U)/\text{Tr}(U_{ideal}^{\dagger}U_{ideal}),
\end{align}
which is utilized to select the optimal evolution trajectory, where $U$ and $U_{ideal}$ represent the error-affected evolution operator and the ideal evolution operator, respectively. We select the controlled-phase gates to validate our proposed scheme
\begin{subequations}
\begin{flalign}
U_{\text{T}}(\gamma')&=\text{diag}(1,1,1,e^{i\gamma'})
\end{flalign}
\end{subequations}
where the basis states are $|00\rangle$, $|01\rangle$, $|10\rangle$ and $|11\rangle$. By varying the parameter $\gamma'$, different controlled phase gates can be achieved. For example, setting $\gamma'$ as $\pi$, $\pi/2$ and $\pi/4$ yield the controlled-Z (CZ), $\text{CP}(\pi/2)$ and $\text{CP}(\pi/4)$ gates, respectively.

As shown in Figs. 1(b) and 1(c), both $\xi_{ZZ}$ and $g_{\text{e}}$ vary with the change of $\omega_{c}-\omega_{1}$. Different evolution trajectories exhibit different sensitivities to ZZ crosstalk. Therefore, to identify the evolution trajectory with optimal robustness against residual crosstalk for the given system parameters, we perform numerical simulations based on the Hamiltonian $H''$ in Eq. (\ref{Hhigh-order}), and plot the results in Fig. \ref{Path}. It is important to note that under the condition $\phi_{\text{AC}}\ll1$, the approximation from the original system Hamiltonian to $H''$ is strictly valid, and the two are essentially consistent. As shown in Fig. \ref{Path}, the optimal evolution trajectory for the $U_{\text{T}}(\gamma')$ gates consistently occurs when $\chi=0.43\pi$. Therefore, once the optimal evolution parameter $\chi$ is determined, $\Delta_{\text{e}}$ and $\phi_{\text{e}}$ for each segment can be respectively determined using Eq. (\ref{pathparameter}), from which $\delta$ and $\varphi$ are subsequently obtained. Finally, based on these parameters, the modulation frequency $\omega_{\phi}$ is calculated according to the formula $\omega_{\phi}=\Delta_{12,\phi}-\alpha_{2}+\delta$. When $g_{\text{e}}/2\pi$ exceeds $2.5 \,\,\text{MHz}$, the stripe patterns emerge in the image as the influence of high-frequency oscillation terms becomes significant. These high-frequency oscillation terms substantially degrade the fidelity, necessitating the selection of an effective coupling strength $g_{\text{e}}/2\pi$ within the range of $1 \,\,\text{MHz}$ to $2 \,\,\text{MHz}$. However, it is important to note that this range of values corresponds to the optimal value only when decoherence errors are not considered. After introducing decoherence errors, the optimal value of the effective coupling strength needs to be reassessed.

Based on the optimal trajectory ($\chi = 0.43\pi$) identified in Fig. \ref{Path}, we further employ the Hamiltonian $H_{e}+\xi_{ZZ}\sigma_{z}^{1}\otimes\sigma_{z}^{2}$ for numerical simulations to evaluate the influence of different crosstalk strengths (corresponding to other system parameters) on fidelity, and thereby deriving the robustness of various schemes, in which $H_{e}$ is the effective Hamiltonian in Eq. (\ref{Heff}), and $\xi_{ZZ}$ represents the ZZ crosstalk strength corresponding to different system parameter settings, ranging from $[-0.1g_{e}, 0.1g_{e}]$. The results are shown in Figs. \ref{Robust}(a)-\ref{Robust}(c), where the schemes we compared are UNGQC scheme $(U_{\text{T}}^{\text{UG}}(\gamma))$, SNGQC scheme $(U_{\text{T}}^{\text{G}}(\gamma))$ and Dynamical scheme $(U_{\text{T}}^{\text{D}}(\gamma))$. The detailed process of the Dynamical scheme can be found in Appendix \ref{APPENDIX A}. To ensure fairness in comparison, all schemes adopt the same effective coupling strength $g_{e}$ in the simulations. These results demonstrate that our work can not only determine the crosstalk robust trajectory for the specific system parameters, but also exhibits strong robustness for other system parameter settings when this evolution trajectory ($\chi=0.43\pi$) is applied.

Simultaneously, since the qubit frequency can be adjusted through the external magnetic flux, this also introduces sensitivity to random flux fluctuations, known as flux noise. Flux noise evidently induces qubit frequency drift. Therefore, it is necessary to further investigate whether our scheme exhibits greater robustness against qubit frequency drift compared to other schemes. Therefore,  to demonstrate the suppression capability of the proposed scheme against qubit frequency drift errors along the evolutionary trajectory ($\chi=0.43\pi$), we plotted Figs. \ref{Robust}(d)-\ref{Robust}(f). The numerical simulations in Figs. \ref{Robust}(d)-\ref{Robust}(f) utilized a Hamiltonian that includes the effective Hamiltonian $H_{e}$ and qubit frequency drift errors in the form of $\omega_{k}\to\omega'_{k} = \omega_{k} + \delta_{k} g_{e}$ with $k=1, 2$. Obviously, our scheme demonstrates a stronger suppression effect against qubit frequency drift.

Therefore, in comparison, our UNGQC scheme demonstrates excellent performance in terms of robustness. Specifically, the suppression of ZZ crosstalk by the triangle-cap trajectory arises from the combination of the intrinsic nature of the geometric phase and trajectory optimization. The magnitude of the geometric phase depends only on the solid angle enclosed by the evolution trajectory in parameter space, and is independent of the specific details of the Hamiltonian evolution. This intrinsic property endows geometric-phase-based quantum gates with a natural immunity to certain types of errors or noise. On this basis, different evolution trajectories exhibit varying sensitivities to different errors. The conventional SNGQC scheme employs a fixed single-loop geometric trajectory, which has limited capability to avoid trajectory segments that are significantly affected by crosstalk.

Moreover, we further conduct trajectory optimization analyses on other system parameter settings. The results demonstrate that our trajectory optimization scheme can stably achieve crosstalk robustness without relying on fixed system parameters, and the trajectory parameter $\chi=0.43\pi$ can serve as a reference value for experimental implementation.

\section{Gate-fidelity evaluation}
Beside, the decoherence induced by the inevitable coupling of quantum system to its surrounding environment is also a consideration in physical implementation. To further analyze the performance of two-qubit gates, we conduct a comprehensive investigation into the impacts of decoherence, high-frequency oscillation error and ZZ crosstalk. We next include these three effects into the quantum dynamics simulated by the Lindblad master equation \cite{master} of 
\begin{align}    \label{master}
\dot{\rho}(t)=-i[{H_{\text{sim}}},\rho(t)]+\sum_{d=-,z}\frac{\kappa_{d}}{2}\mathcal L (D_{d}),
\end{align}
where $\rho$ is density operator, and $d=-,z$ is to distinguish decay and dephasing operator, respectively. $\kappa_{d}$ represent the decay or dephasing rate. $\mathcal L(D_{d})=2D_{d}\rho D_{d}^{\dagger}-D_{d}^{\dagger}D_{d}\rho-\rho D_{d}^{\dagger}D_{d}$ is the Lindblad operator. In simulation, we consider the decay and dephasing operators of $D_{-}=|0\rangle\langle1|+\sqrt{2}\;|1\rangle\langle2|$, $D_{z}=2\,|2\rangle\langle2|+|1\rangle\langle1|$. Therefore, when $H_{\text{sim}}=H''$, we can determine the evolution of the density operator under the influence of decoherence, high-frequency oscillation error and ZZ crosstalk by solving the master equation. Following the definition of gate fidelity, we can average the fidelity across all initial states to ultimately derive the gate fidelity \cite{F4, F5}
\begin{align}     \label{Fdec}
F_{\text{dec}}(t)=\dfrac{1}{4\pi^{2}}\int_{0}^{2\pi}\int_{0}^{2\pi}\langle\psi(\tau)|\rho(t)|\psi(\tau)\rangle d\theta_{1} d\theta_{2},
\end{align}
where $|\psi(\tau)\rangle=U(\tau)|\psi_{0}\rangle$ is the ideal final state with a general initial state being $|\psi_{0}\rangle = (\cos{\theta_{1}}|0\rangle_{1}+\sin{\theta_{1}}|1\rangle_{1}) \otimes (\cos{\theta_{2}}|0\rangle_{2}+\sin{\theta_{2}}|1\rangle_{2})$. Therefore, even before evolution begins (i.e., at $t=0$), some initial states are not completely orthogonal to certain ideal final states, resulting in an average state fidelity (i.e., gate fidelity) greater than zero. Hence, in this context, the simulated gate fidelity corresponds to the average state fidelity over a set of initial states after they have undergone the gate operation, which is clearly different from the fidelity trend of a specific state, particularly at the initial time. Consequently, the starting point of gate fidelity varies depending on the specific gate operation. 

To exclude the influence of other error sources (such as high-frequency oscillations) and evaluate the sensitivity of each scheme to decoherence, we use the effective Hamiltonian $H_{\text{sim}}=H_{\text{e}}$ and incorporate decoherence effects (including energy decay and dephasing) via the master equation for numerical simulations, resulting in the plot shown in Figs. \ref{Decoherence}(a)-\ref{Decoherence}(c), where $\kappa=\kappa_{-}=\kappa_{z}$. The simulation results clearly demonstrate that our UNGQC scheme has the lowest sensitivity to qubit decoherence. Specifically, when $\kappa=g_{e}/500$, the fidelity of $U_{\text{T}}(\pi)$ gate in the UNGQC scheme improves by $0.7\,\%$ and $1.4\,\%$ respectively compared to the SNGQC scheme and Dynamical scheme. Other gate types are also improved accordingly.

To evaluate the gate fidelity under realistic superconducting implementation conditions, including decoherence, high-frequency oscillation terms, and ZZ crosstalk, we consider the Hamiltonian $H_{\text{sim}}=H''$ and incorporate decoherence effects via the master equation (with a uniform decoherence rate of $\kappa=\kappa_{-}=\kappa_{z}=2\pi\,\times \, 2\;\text{kHz}$), thus illustrating the results through numerical simulation in Figs. \ref{Decoherence}(d)-\ref{Decoherence}(f). Unlike the previous case, after considering the effects of decoherence, high-frequency oscillation terms, and ZZ crosstalk, the optimal value of $g_{\text{e}}$ is approximately $2\pi \times 4.3\,\, \text{MHz}$. It is necessary to emphasize again that under the weak-driving condition $\phi_{\text{AC}}\ll1$, the approximation from the full system Hamiltonian (i.e., Eq. (\ref{Hyuan})) to $H''$ is rigorously valid, and the difference in fidelities obtained from simulating the two is less than $0.01\%$, confirming the validity of this approximation and allowing for the simulation of errors in actual physical processes. Furthermore, due to the fact that different segments correspond to different effective Hamiltonians, the system parameters will experience discontinuous changes at the moments when the effective Hamiltonian undergoes transformations (i.e., at the moments $\tau_{1}$ and $\tau_{2}$ in the UNGQC scheme, or at the moment $\tau_{1}$ in the SNGQC scheme), causing the gate fidelity to exhibit discontinuous variations during the time evolution process. The final simulation results demonstrate that, under realistic conditions, our UNGQC scheme still achieves high gate fidelity. Currently, multiple research groups have experimentally realized geometric quantum gates and achieved high fidelity, further demonstrating the feasibility of our approach \cite{SGR2,SGR3}.

\section{Conclusion}

In conclusion, we have proposed a coupler-assisted superconducting two-qubit geometric gate scheme that effectively addresses the trade-off between crosstalk suppression and gate duration. By leveraging additional parametric degrees of freedom to steer the evolution away from crosstalk-sensitive regions, the scheme enables fast and high-fidelity two-qubit operations. Numerical results confirm its strong robustness against  typical experimental imperfections such as qubit frequency drift, and its ability to maintain high performance even under realistic conditions including high-frequency oscillations and decoherence.

\acknowledgments

This work was supported by the National Natural Science Foundation of China (Grants No. 12305019, No. 92576110, and No.  12275090), and the Guangdong Provincial Quantum Science Strategic Initiative (Grant No. GDZX2203001).

\appendix

\section{DYNAMICAL SCHEME} \label{APPENDIX A}

In general, the dynamical gates utilize a resonant two-level Hamiltonian with a constant phase parameter. Therefore, for the effective Hamiltonian within the $\{|11\rangle, |02\rangle\}$ subspace shown in Eq. (11), we set the effective Hamiltonian parameters as $\Delta_{\text{e}}=0$, $\phi_{\text{e}}(t)=\phi_{\text{D}}$, thus
\begin{align}
H_{\text{D}}(t)=\frac{1}{2} \begin{pmatrix}
0 & g_{\text{e}}e^{-i\phi_{\text{D}}}\\
g_{\text{e}}e^{i\phi_{\text{D}}} & 0
\end{pmatrix}.
\end{align}
Correspondingly, the evolution operator  at the final moment $\tau$ can be written as:
\begin{align}
U_{\text{D}}(\theta_{\text{D}},\phi_{\text{D}})&=e^{-i\int_{0}^{\tau}H_{\text{D}}(t)dt}\notag\\
&=\begin{pmatrix}
\cos{\frac{\theta_{\text{D}}}{2}} & -ie^{-i\phi_{\text{D}}}\sin{\frac{\theta_{\text{D}}}{2}}\\
-ie^{i\phi_{\text{D}}}\sin{\frac{\theta_{\text{D}}}{2}} & \cos{\frac{\theta_{\text{D}}}{2}}
\end{pmatrix},
\end{align}
where $\theta_{\text{D}}=\int_{0}^{\tau}g_{\text{e}}(t)dt$. By setting different values of $\theta_{\text{D}}$ and $\phi_{\text{D}}$ , various two-qubit gates can be obtained. To realize the gate $U_{\text{T}}(\gamma')=\text{diag}(1,1,1,e^{i\gamma'})$ within the dynamical scheme as well, we need to specify a gate sequence:
\begin{align}
U_{\text{T}}^{\text{D}}(\theta_{z})=U_{\text{D}}(\frac{\pi}{2},\pi)U_{\text{D}}(\theta_{z},-\frac{\pi}{2})U_{\text{D}}(\frac{\pi}{2},0).
\end{align}
In this process, the geometric phase $\gamma_{g}$ accumulated by the quantum state is zero. Thus, the two-qubit gate constructed in this way is called a dynamical gate. Accordingly, $U_{\text{T}}^{\text{D}}(\pi)$ corresponds to the CZ gate, $U_{\text{T}}^{\text{D}}(\pi/2)$ corresponds to the CP($\pi/2$) gate, and $U_{\text{T}}^{\text{D}}(\pi/4)$ corresponds to the CP($\pi/4$) gate.


\begin{thebibliography}{99}

\bibitem{QC}
M. A. Nielsen and I. L. Chuang, \textit{Quantum Computation and Quantum Information} (Cambridge University Press, 2000).

\bibitem{ZP}
P. Zhao, P. Xu, D. Lan, J. Chu, X. Tan, H.-F. Yu, and Y. Yu, High-contrast ZZ interaction using superconducting qubits with opposite-sign anharmonicity, Phys. Rev. Lett. \textbf{125}, 200503 (2020).

\bibitem{ZZ and geff}
X.-Y. Han, T.-Q. Cai, X.-G. Li, Y.-K. Wu, Y.-W. Ma, and Y.-L. Ma, Error analysis in suppression of unwanted qubit interactions for a parametric gate in a tunable superconducting circuit, Phys. Rev. A \textbf{102}, 022619 (2020).

\bibitem{two qubit}
M. Kjaergaard, M. E. Schwartz, J. Braum\"{u}ller, P. Krantz, J. I.-J. Wang, S. Gustavsson, and W. D. Oliver, Superconducting qubits: current state of play, Annu. Rev. Condens. Matter Phys. \textbf{11}, 369 (2020).

\bibitem{cross-resonance1}
J. M. Chow, A.D. C\'{o}rcoles, J. M. Gambetta, C. Rigetti, B. R. Johnson, J. A. Smolin, J. R. Rozen, G. A. Keefe, M. B. Rothwell, M. B. Ketchen, and M. Steffen, Simple all-microwave entangling gate for fixed-frequency superconducting qubits, Phys. Rev. Lett. \textbf{107}, 080502 (2011).

\bibitem{cross-resonance2}
C. Rigetti and M. Devoret, Fully microwave-tunable universal gates in superconducting qubits with linear couplings and fixed transition frequencies, Phys. Rev. B \textbf{81}, 134507 (2010).

\bibitem{cross-resonance3}
J. C. Pommerening, Multiqubit coupling dynamics and the cross-resonance gate, Ph.D. thesis, RWTH Aachen University, 2017.

\bibitem{cross-resonance4}
S. Sheldon, E. Magesan, J. M. Chow, and J. M. Gambetta, Procedure for systematically tuning up cross-talk in the cross-resonance gate, Phys. Rev. A \textbf{93}, 060302(R) (2016).

\bibitem{cross-resonance5}
M. E. Ware, Flux-tunable superconducting transmons for quantum information processing, Ph.D. thesis, University of Alabama, 2009.

\bibitem{cross-resonance6}
S. Kirchhoff, T. Ke{\ss}ler, P. J. Liebermann, E. Ass\'{e}mat, S. Machnes, F. Motzoi, and F. K. Wilhelm, Optimized cross resonance gate for coupled transmon systems, Phys. Rev. A \textbf{97}, 042348 (2018).

\bibitem{parametrically tunable coupling0}
A. O. Niskanen, Y. Nakamura, and J. S. Tsai, Tunable coupling scheme for flux qubits at the optimal point, Phys. Rev. B \textbf{73}, 094506 (2006).

\bibitem{parametrically tunable coupling1}
A. O. Niskanen, K. Harrabi, F. Yoshihara, Y. Nakamura, S. Lloyd, and J. S. Tsai, Quantum coherent tunable coupling of superconducting qubits, Science \textbf{316}, 723 (2007).

\bibitem{parametrically tunable coupling2}
J. D. Strand, M. Ware, F. Beaudoin, T. A. Ohki, B. R. Johnson, A. Blais, and B. L. T. Plourde, First-order sideband transitions with flux-driven asymmetric transmon qubits, Phys. Rev. B \textbf{87}, 220505 (2013).

\bibitem{parametrically tunable coupling3}
D. C. McKay, S. Filipp, A. Mezzacapo, E. Magesan, J. M. Chow, and J. M. Gambetta, Universal gate for fixed-frequency qubits via a tunable bus, Phys. Rev. Appl. \textbf{6}, 064007 (2016).

\bibitem{parametrically tunable coupling4}
Y. Lu, S. Chakram, N. Leung, N. Earnest, R. K. Naik, Z. Huang, P. Groszkowski, E. Kapit, J. Koch, and D. I. Schuster, Universal stabilization of a parametrically coupled qubit, Phys. Rev. Lett. \textbf{119}, 150502 (2017).

\bibitem{parametrically tunable coupling5}
R. K. Naik, N. Leung, S. Chakram, P. Groszkowski, Y. Lu, N. Earnest, D. C. McKay, J. Koch, and D. I. Schuster, Randomaccess quantum information processors using multimode circuit Quantum Electrodynamics, Nat. Commun. \textbf{8}, 1715 (2017).

\bibitem{parametrically tunable coupling6}
M. Roth, M. Ganzhorn, N. Moll, S. Filipp, G. Salis, and S. Schmidt, Analysis of a parametrically driven exchange-type gate and a two-photon excitation gate between superconducting qubits, Phys. Rev. A \textbf{96}, 062323 (2017).

\bibitem{parametrically tunable coupling7}
M. Reagor, C. B. Osborn, N. Tezak, A. Staley, G. Prawiroatmodjo, M. Scheer, N. Alidoust, E. A. Sete, N. Didier, {\it et al}., Demonstration of universal parametric entangling gates on a multi-qubit lattice, Sci. Adv. \textbf{4}, eaao3603 (2018).

\bibitem{parametrically tunable coupling8}
S. Caldwell, N. Didier, C. A. Ryan, E. A. Sete, A. Hudson, P.Karalekas, R. Manenti, M. Reagor, M. P. da Silva, R. Sinclair, E. Acala, {\it et al}., Parametrically activated entangling gates using transmon qubits, Phys. Rev. Appl. \textbf{10}, 034050 (2018).

\bibitem{parametrically tunable coupling9}
X. Li, Y. Ma, J. Han, T. Chen, Y. Xu, W. Cai, H. Wang, Y. P. Song, Z.-Y. Xue, Z.-Q. Yin, and L. Sun, Perfect quantum state transfer in a superconducting qubit chain with parametrically tunable couplings, Phys. Rev. Appl. \textbf{10}, 054009 (2018).

\bibitem{YAN}
F. Yan, P. Krantz, Y. Sung, M. Kjaergaard, D. L. Campbell, T. P. Orlando, S. Gustavsson, and W. D. Oliver, Tunable coupler scheme for implementing high-fidelity two-qubit gates, Phys. Rev. Appl. \textbf{10}, 054062 (2018).

\bibitem{tunable coupler2}
E. A. Sete, N. Didier, A. Q. Chen, S. Kulshreshtha, R. Manenti, and S. Poletto, Parametric-resonance entangling gates with a tunable coupler, Phys. Rev. Appl. \textbf{16}, 024050 (2021).

\bibitem{tunable coupler3}
T.-Q. Cai, X.-Y. Han, Y.-K. Wu, Y.-L. Ma, J.-H. Wang, Z.-L. Wang, H.-Y Zhang, H.-Y Wang, Y.-P. Song, and L.-M. Duan, Impact of spectators on a two-qubit gate in a tunable coupling superconducting circuit, Phys. Rev. Lett. \textbf{127}, 060505 (2021).

\bibitem{tunable coupler4}
Y. Sung, L. Ding, J. Braum\"{u}ller, A. Veps\"{a}l\"{a}inen, B. Kannan, M. Kjaergaard, A. Greene, G. O. Samach, C. McNally, D. Kim, A. Melville, B. M. Niedzielski, M. E. Schwartz, J. L. Yoder, T. P. Orlando, S. Gustavsson, and W. D. Oliver, Realization of high-fidelity CZ and ZZ-Free iSWAP gates with a tunable coupler, Phys. Rev. X \textbf{11}, 021058 (2021).

\bibitem{tunable coupler5}
K. Luo, W.-H. Huang, Z.-Y. Tao, L.-B. Zhang, Y.-X. Zhou, J. Chu, W.-X. Liu, B.-Y. Wang, J.-Y Cui, S. Liu, F. Yan, M.-H. Yung, Y. Chen, T. Yan, and D. Yu, Experimental realization of two qutrits gate with tunable coupling in superconducting circuits, Phys. Rev. Lett. \textbf{130}, 030603 (2023).

\bibitem{tunable coupler6}
S. Li, D. Fan, M. Gong, X. Chen, Y. Wu, H. Guan, H. Deng, H. Rong, H.-L. Huang, \textit{et al}., Realization of fast all-microwave controlled-Z gates with a tunable coupler, Chin. Phys. Lett. \textbf{39}, 030302 (2022).

\bibitem{crosstalk review 1}
Y. Y. Gao, M. A. Rol, S. Touzard, and C. Wang, Practical guide for building superconducting quantum devices, PRX Quantum \textbf{2}, 040202 (2021).

\bibitem{crosstalk review 2}
Z. Zhou, R. S., Y. Oda , K. Schultz , and G. Quiroz, Quantum crosstalk robust quantum control, Phys. Rev. Lett. \textbf{131}, 210802 (2023).

\bibitem{CT1 1}
A. Kandala, K. X. Wei, S. Srinivasan, E. Magesan, S. Carnevale, G. A. Keefe, D. Klaus, O. Dial, and D. C. McKay, Demonstration of a high-fidelity CNOT gate for fixed-frequency transmons with engineered  ZZ suppression, Phys. Rev. Lett. \textbf{127}, 130501 (2021).

\bibitem{CT1 2}
P. Mundada, G. Zhang, T. Hazard, and A. Houck, Suppression of qubit crosstalk in a tunable coupling superconducting circuit, Phys. Rev. Appl. \textbf{12}, 054023 (2019).

\bibitem{CT2 1}
P. Zhao, K. Linghu, Z. Li, P. Xu, R. Wang, G. Xue, Y. Jin, and H. Yu, Quantum crosstalk analysis for simultaneous gate operations on superconducting qubits, PRX Quantum \textbf{3}, 020301 (2022).

\bibitem{CT2 2}
S. Krinner, S. Lazar, A. Remm, C.K. Andersen, N. Lacroix, G.J. Norris, C. Hellings, M. Gabureac, C. Eichler, and A. Wallraff, Benchmarking coherent errors in controlled-phase gates due to spectator qubits, Phys. Rev. Applied \textbf{14}, 024042 (2020).

\bibitem{CT3 1}
S. Kosen, H.-X. Li, M. Rommel, R. Rehammar, M. Caputo, L. Gr\"{o}nberg, J. Fern\'{a}ndez-Pend\'{a}s, A. F. Kockum, J. Bizn\'{a}rov\'{a}, L. Chen, C. Kri\v{z}an, A. Nylander, A. Osman, A. F. Roudsari, D. Shiri, G. Tancredi, J. Govenius, and J. Bylander, Signal crosstalk in a flip-chip quantum processor, PRX Quantum \textbf{5}, 030350 (2024).

\bibitem{ZZ reason1}
Z. Ni, S. Li, L. Zhang, J. Chu, J. Niu, T. Yan, X. Deng, L. Hu, J. Li, Y. Zhong, S. Liu, F. Yan, Y. Xu, and D. Yu, Scalable method for eliminating residual ZZ interaction between superconducting qubits, Phys. Rev. Lett. \textbf{129}, 040502 (2022).

\bibitem{ZZ reason2}
X. Li, T. Cai, H. Yan, Z. Wang, X. Pan, Y. Ma, W. Cai, J. Han, Z. Hua, X. Han, Y. Wu, H. Zhang, H. Wang, Y. Song, L. Duan, and L. Sun, Tunable coupler for realizing a controlled-phase gate with dynamically decoupled regime in a superconducting circuit, Phys. Rev. Appl. \textbf{14}, 024070 (2022).

\bibitem{DYN1}
V. Tripathi, H. Chen, M. Khezri, K. Yip, E.M. Levenson-Falk, and D. A. Lidar, Suppression of crosstalk in superconducting qubits using dynamical decoupling, Phys. Rev. Appl. \textbf{18}, 024068 (2022).

\bibitem{DYN2}
S.-Y. Niu, A. Todri-Sanial, and N. T. Bronn, Multi-qubit dynamical decoupling for enhanced crosstalk suppression, Quantum Sci. Technol. \textbf{9}, 045003 (2024).

\bibitem{DYN3}
K. Nakamura and J. Ankerhold, Impact of time-retarded noise on dynamical decoupling schemes for qubits, Phys. Rev. B \textbf{111}, 064503 (2025).

\bibitem{DYN4}
A. F. Brown and D. A. Lidar, Efficient chromatic-number-based multiqubit decoherence and crosstalk suppression, PRX Quantum \textbf{6}, 020354 (2025).

\bibitem{DYN5}
A. Aaliray and H. Mohammadi, Quantum speed limit time in two-qubit system by dynamical decoupling method, Sci. Rep. \textbf{15}, 8338 (2025).

\bibitem{DYN6}
H. Jeong, Y. Kim, B. Choi, M. Cho, S. Woo, Y. Chong, Y. Lee, and H. Yeo, Suppressing spectator-induced dephasing through optimized dynamical decoupling implementation, Sci. Rep. \textbf{15}, 18698 (2025).

\bibitem{LIANG YAN}
Y. Liang, M.-J. Liang, S. Li, Z. D. Wang, and Z.-Y. Xue, Scalable protocol to mitigate ZZ crosstalk in universal quantum gates, Phys. Rev. Appl. \textbf{21}, 024016 (2024).

\bibitem{SW1}
A. Blais, J. Gambetta, A. Wallraff, D. I. Schuster, S. M. Girvin, M. H. Devoret, and R. J. Schoelkopf, Quantum information processing with circuit quantum electrodynamics, Phys. Rev. A \textbf{75}, 032329 (2007).

\bibitem{SW2}
S. Bravyi, D. P. DiVincenzo, and D. Loss, Schrieffer–wolff transformation for quantum many-body systems, Ann. Phys. (N.Y.) \textbf{326}, 2793 (2011).

\bibitem{omega c}
J. Koch, T. M. Yu, J. Gambetta, A. A. Houck, D. I. Schuster, J. Majer, A. Blais, M. H. Devoret, S. M. Girvin, and R. J. Schoelkopf, Charge-insensitive qubit design derived from the Cooper pair box, Phys. Rev. A \textbf{76}, 042319 (2007).

\bibitem{P2}
H.-L. Huang, D. Wu, D. Fan, and X. Zhu, Superconducting quantum computing: A review, Sci. China Inf. Sci. \textbf{63}, 180501 (2020).

\bibitem{P3}
C. Wang, X. Li, H. Xu, Z. Li, J. Wang, Z. Yang, Z. Mi, X. Liang, T. Su, C. Yang, \textit{et al}., Towards practical quantum computers: Transmon qubit with a lifetime approaching 0.5 ms, npj Quantum Inf. \textbf{8}, 1 (2022).

\bibitem{perturbation}
G. Zhu, D. G. Ferguson, V. E. Manucharyan, and J. Koch, Circuit QED with fluxonium qubits: Theory of the dispersive regime, Phys. Rev. B \textbf{87}, 024510 (2013).

\bibitem{geometric phase}
Y. Aharonov and J. Anandan, Phase change during a cyclic quantum evolution, Phys. Rev. Lett. \textbf{58}, 1593 (1987).

\bibitem{wang2001} X.-B. Wang and M. Keiji, Nonadiabatic conditional geometric
phase shift with NMR, Phys. Rev. Lett. {\bf 87}, 097901 (2001).

\bibitem{zhu2002}
S. L. Zhu and Z. D. Wang, Implementation of Universal Quantum Gates Based on Nonadiabatic Geometric Phases, Phys. Rev. Lett. {\bf 89}, 097902 (2002).



\bibitem{xuereview}
Z.-Y. Xue and C.-Y. Ding, Recent advances on nonadiabatic geometric quantum computation, Front. Phys. 21, 033202 (2026).

\bibitem{SNGQC3}
P. Z. Zhao, X.-D. Cui, G. F. Xu, E. Sj\"oqvist, and D. M. Tong, Rydberg-atom-based scheme of nonadiabatic geometric quantum computation, Phys. Rev. A \textbf{96}, 052316 (2017).


\bibitem{SNGQC1}
T. Chen and Z.-Y. Xue, Nonadiabatic geometric quantum computation with parametrically tunable coupling, Phys. Rev. Appl. \textbf{10}, 054051 (2018).


\bibitem{liangyan}
Y. Liang and Z.-Y. Xue,
Nonadiabatic geometric quantum gates with on-demand trajectories, Phys. Rev. Appl. \textbf{21}, 064048 (2024).


\bibitem{SNGQC2}
C. Scarato, K. Hanke, A. Remm, S. Laz\v{a}r, N. Lacroix, D. C. Zanuz, A. Flasby, A. Wallraff, and C. Hellings, Realizing a continuous set of two-qubit gates parameterized by an idle time, PRX Quantum \textbf{6}, 040317 (2025).

\bibitem{F1}
A. Gilchrist, N. K. Langford, and M. A. Nielsen, Distance measures to compare real and ideal quantum processes, Phys. Rev. A \textbf{71}, 062310 (2005).

\bibitem{F2}
L. H. Pedersen, N. M. M{\o}ller, and K. M{\o}lmer, Fidelity of quantum operations, Phys. Lett. A \textbf{367}, 47 (2005).

\bibitem{F3}
X. Wang, Z. Sun, and Z. D. Wang, Operator fidelity susceptibility: An indicator of quantum criticality, Phys. Rev. A \textbf{79}, 012105 (2009).

\bibitem{master}
F. Motzoi, J. M. Gambetta, P. Rebentrost, and F. K. Wilhelm, Simple pulses for elimination of leakage in weakly nonlinear qubits, Phys. Rev. Lett. \textbf{103}, 110501 (2009).

\bibitem{F4}
J. F. Poyatos, J. I. Cirac, and P. Zoller, Complete characterization of a quantum process: The two-bit quantum gate, Phys. Rev. Lett. \textbf{78}, 390 (1997).

\bibitem{F5}
Z.-Q. Yin and F.-L. Li, Multiatom and resonant interaction scheme for quantum state transfer and logical gates between two remote cavities via an optical fiber, Phys. Rev. A \textbf{75}, 012324 (2007).

\bibitem{SGR2}
K. Xu, W. Ning, X.-J. Huang, P.-R. Han, H. Li, Z.-B. Yang, D. Zheng, H. Fan, and S.-B. Zheng, Demonstration of a non-Abelian geometric controlled-NOT gate in a superconducting circuit, Optical \textbf{8}, 972 (2021).

\bibitem{SGR3}
Y. Xu, Z. Hua, T. Chen, X. Pan, X. Li, J. Han, W. Cai, Y. Ma, H. Wang,  Y. P. Song, Z.-Y. Xue, and L. Sun, Experimental implementation of universal nonadiabatic geometric quantum gates in a superconducting circuit, Phys. Rev. Lett. \textbf{124}, 230503 (2020).

\end{thebibliography}
\end{document}